\documentclass{elsart}

\usepackage{graphics}
\usepackage[pdftex]{graphicx}

\usepackage{amssymb}
\usepackage{amsmath}
\usepackage{rotating}
\usepackage{epsfig}
\usepackage{subfigure}
\usepackage{boxedminipage}
\usepackage{enumerate}

\begin{document}

\begin{frontmatter}



\title{Billion-atom Synchronous Parallel Kinetic Monte Carlo Simulations of Critical 3D Ising Systems}




\author[label1]{E. Mart\'inez},
\author[label2]{P. R. Monasterio}, and
\author[label3]{J. Marian}

\address[label1]{IMDEA-Materiales, Madrid 28040, Spain}
\address[label2]{Massachusetts Institute of Technology, Cambridge, MA 02139}
\address[label3]{Lawrence Livermore National Laboratory, Livermore, CA 94551}

\begin{abstract}
An extension of the synchronous parallel kinetic Monte Carlo (pkMC) algorithm developed by 
Martinez {\it et al} [{\it J.\ Comp.\ Phys.}~{\bf 227} (2008) 3804] to discrete lattices 
is presented. The method solves the master equation synchronously by recourse to null events 
that keep all processors time clocks current in a global sense. Boundary conflicts are rigorously 
solved by adopting a chessboard decomposition into non-interacting sublattices. We find that 
the bias introduced by the spatial correlations attendant to the sublattice decomposition is  
within the standard deviation of the serial method, which confirms the statistical validity 
of the method. We have assessed the parallel efficiency of the method and find that our algorithm 
scales consistently with problem size and sublattice partition. We apply the method to the calculation 
of scale-dependent critical exponents in billion-atom 3D Ising systems, with very good agreement with state-of-the-art 
multispin simulations. 
\end{abstract}

\end{frontmatter}

\section{Introduction}
\label{sec:intro}
Kinetic Monte Carlo (kMC) \cite{kalosbook} has proven an efficient and powerful tool to study non-equilibrium 
processes, and is used in fields as different as population dynamics, irradiation damage, or crystal growth 
\cite{fichthorn1991,voter2006}. 
The most widely used variant of the method is the Monte Carlo \emph{time residence} algorithm \cite{cox1965}, 
also known as rejection-free $n$-fold method, or BKL in reference to its authors \cite{bkl}. Although kMC is 
generally capable of advancing the time scale significantly faster than direct, time-driven methods, it suffers 
from numerical limitations such as \emph{stiffness} \cite{snyder2005}, and time asynchronicity. This has spurred 
the development of more powerful variants such as coarse-grained kMC \cite{chatterjee2005}, first-passage kMC 
\cite{oppelstrup}, and other accelerated methods \cite{chatterjee2007}. 
In this sense, a number of parallelization schemes for kMC have been proposed, including rigorous 
and semi-rigorous algorithms based on asynchronous kinetics \cite{lubachevsky1988,eick1993,shim2006}. 
These methods rely on cumbersome roll-back procedures to avoid causality errors, {\it i.e.} event time 
incompatibilities associated with processor communications. For this reason, most applications of interest 
are studied using approximate schemes (non-rigorous) for computational convenience. In spite of this, 
calculations using asynchronous parallel kMC have provided numerous insights in several studies, most
notably crystal growth \cite{nandipati2009}. 

Recently, we have developed and alternative algorithm based on a synchronous time decomposition of the 
master equation \cite{martinez2008}. Our parallel kinetic Monte Carlo method, eliminates time 
conflicts by recourse to \emph{null} events that advance the internal clock of each processor in a synchronized 
fashion without altering the stochastic trajectory of the system. The method has been demonstrated for 
continuum diffusion/reaction systems, which represents a worst-case application scenario for two reasons. 
First, the maximum time step gain is limited by the intrinsic length scale of the problem at hand, which 
in concentrated systems may not be large; second, spatial boundary errors are difficult to eliminate due 
to the unbounded nature of diffusion in a continuum setting. This latter feature also limits the parallel 
efficiency of the algorithm, as global communications are needed during every Monte Carlo step. In our synchronous 
parallel kMC method (spkMC), the parallel error can always be computed intrinsically and reduced arbitrarily 
(albeit at the expense of efficiency). 
In this paper, we extend spkMC to lattice systems, where diffusion lengths are quantized and boundary errors 
can be eliminated altogether. First, we adapt the algorithm proposed in Ref.\ \cite{martinez2008} to discrete 
systems. Due to its relevance and well-known properties, we have chosen the three-dimensional (3D) 
Ising system as our underlying lattice model. Second, we analyze the performance of the method 
in terms of stochastic bias (error) and parallel efficiency. 
We then apply spkMC to large systems near the critical point, which provides a demanding testbed 
for the method, as this is where fluctuations are exacerbated and convergence is most difficult.

\section{The 3D Ising model}
\label{sec:Ising}
The Ising model is one of the most extensively studied lattice systems in statistical mechanics. 
It consists of a lattice of $N$ sites occupied by particles whose state is described by a variable 
$\sigma_i$ that represents the spin of each particle and can take only the value $\pm1$. 
For pair-interactions, the Hamiltonian that gives the energy of 
configurational state $\sigma=\{\sigma_i\}$ takes the form:
\begin{equation}
	\mathcal{H}(\sigma)=-J\sum_{\langle i,j\rangle}\sigma_i\sigma_j-H\sum_i\sigma_i
	\label{eq:isingh}
\end{equation}
where $J$ is the coupling constant between neighboring pairs $\langle i,j\rangle$ and $H$ is an 
external (magnetic) field. The time evolution of the system is assumed to be described by a master 
equation with Glauber dynamics \cite{glauber1963}, which states that the probability $p(\sigma,t)$ 
of finding the system in state $\sigma$ at time $t$ obeys the equation:
\begin{equation}
	\frac{\partial p(\sigma,t)}{\partial t}=\sum_i\left[ \mathcal{W}_i(\sigma)p(\sigma,t) -  \mathcal{W}_i(\sigma')p(\sigma',t)\right]
	\label{eq:glauber}
\end{equation}
where $\sigma'$ denotes the configuration obtained from $\sigma$ by flipping the $i^{\mathrm{th}}$ 
spin with transition rate $\mathcal{W}_i(\sigma)$:
\begin{equation}
	\mathcal{W}_i(\sigma)=\frac{\lambda}{2}\left[1-\sigma_i\tanh\left(2\beta\Delta E_i\right)\right]
	\label{eq:rate}
\end{equation}
Here $\lambda$ is a positive constant that represents the natural frequency of the system, 
$\beta$ is the reciprocal temperature, and $\Delta E_i=-J\sum_{\langle i,j\rangle}\sigma_j-H\sigma_i$ 
is the energy associated with spin $i$, which follows directly from eq.\ (\ref{eq:isingh}).
In what follows we consider only internally-driven systems ($H=0$).

Many discrete systems can be mapped exactly or approximately to the Ising system. 
The grand canonical ensemble formulation of the lattice gas model, for example, 
can be mapped exactly to the canonical ensemble formulation of the Ising model. 
Also, binary alloy hamiltonians with nearest-neighbor interactions in rigid lattices 
can also be expressed as Ising hamiltonians. These mappings allow us to exploit results
and behaviors of the Ising model to answer questions about the related models. In addition, 
the Ising system is particularly useful to study second-order phase transitions. 
The temperature $T_c$ at which such transitions occur is known as the critical temperature. 
During the phase transition, thermodynamic quantities diverge according to power laws of 
$T$, whose exponents are known as the critical exponents. The nature of the phase transition 
is determined by whether the order parameter is continuous at $T_c$ \cite{newman}. 
In a ferromagnetic system such as the Ising model, the order parameter is the net magnetization $m(\sigma)$:
\begin{equation}
	m(\sigma) = \frac{1}{N}\sum_i\sigma_i
	\label{eq:magnet}
\end{equation}
For simple cubic lattices in 3D, $N=L^3$ is the number of sites, with $L$ the lattice size. 
As we approach the critical temperature from $T>T_c$, uncorrelated groups of spins 
align themselves in the same direction. These clusters grow in size, known as the 
correlation length $\xi$, which too diverges at the critical point. 
At $T=T_c$, one may theoretically encounter arbitrarily large areas with correlated 
spins pointing in one direction. In finite systems the upper limit of $\xi$ is the 
system's dimension $L$. Thus, the challenge associated with simulating Ising systems 
during the phase transition is then ensuring that the error incurred by simulating a 
finite-size lattice is sufficiently small for the critical exponents to be calculated 
with certainty. This has spurred a great many Monte Carlo simulations of very large 
lattices in the hope of finding converged critical exponents (cf., {\it e.g.}, Ref.\ \cite{hasenbusch2001}).

When the system is in a ferromagnetic state, $m$ decays from the spontaneous 
magnetization value $m_0$ with time as $m\propto t^{-\kappa/\nu z}$, 
where $\kappa$ and $\nu$ are the critical exponents for $m_0$ and $\xi$, respectively. 
From the known value of the ratio $\kappa/\nu=0.515$ \cite{berche2004} in 3D, 
one can obtain $z$ from the 
slope of the $m$-$t$ curve, obtained for several $L$, at the critical point. 
To study the finite-size dependence of $T_c$, high-order dimensionless ratios 
such as the \emph{Binder} cumulant have been proposed:
\begin{equation}
	U_4 = \frac{\langle m^4\rangle}{\langle m^2\rangle^2}
	\label{eq:binder}
\end{equation}
which takes a value of $U_4\approx3$ when $T>T_c$ 
(when the magnetization oscillates aggressively around zero), 
and goes to zero at low temperatures, when $m=m_0$. 
As mentioned earlier, at the critical point the correlation 
length diverges, and therefore $U_4$ does not depend on $L$.  
kMC equilibrium calculations of $U_4$ for several lattice sizes can then 
be used to calculate the value of the reciprocal critical temperature
$\beta_c=J/kT_c$, whose most accurate estimate is presently 
$\beta_c=0.2216546$ \cite{baillie1992,novotny2003}.

\section{Parallel algorithm for lattice systems}
\label{sec:algo}

\subsection{General algorithm}
\label{subsec:alg}
The basic structure of the algorithm is identical to that described in Ref.\ \cite{martinez2008}. 
First, the entire configurational space is partitioned into $K$ subdomains $\Omega_k$. Note that, 
in principle, this decomposition need not be necessarily spatial (although this is the most common one), and
partitions based on some other kind of load balancing can be equally adopted. However, without loss of 
generality, in what follows it is assumed that the system is spatially partitioned:
\begin{enumerate}
	\item[\bf 1.] {\bf A frequency line is constructed for each $\Omega_k$ as the aggregate of the individual rates, $r_{ik}$, corresponding to all the possible events within each subdomain:
	\begin{displaymath}
    	\label{eq:step1}
    		R_k = \sum_i^{n_k} r_{ik}
	\end{displaymath}}
where $n_k$ and $R_k$ are, respectively, the number of possible events and the total rate in each subdomain $k$. Here $R_{tot} = \sum_k^K R_{k}$ and $N=\sum_k^K n_k$.
	\item[\bf 2.] {\bf We define the maximum rate, $R_{max}$, as:
	\begin{displaymath}
    	\label{eq:step2}
    		R_{max}\ge\max_{k=1,...K}{\{R_k\}}
	\end{displaymath}}
This value is then communicated globally to all processors.
	\item[\bf 3.] {\bf We assign a \emph{null} event with rate $r_{0k}$ to each frequency line in each subdomain $k$ such that:
	\begin{displaymath}
	\label{eq:step3}
    		r_{0k} = R_{max} - R_k
	\end{displaymath}}
where, in general, the $r_{0k}$ will all be different. 
We showed in Ref.\ \cite{martinez2008} that the condition for maximum efficiency is 
that step (2) become strictly an equality, such that:
	\begin{displaymath}
		\exists~\Omega_{\alpha},~\alpha\in\{k\},~|~R_{\alpha}\equiv R_{max}\rightarrow r_{\alpha0}=0
	\end{displaymath}
{\it i.e.} there is no possibility of null events. However, in principle, 
each subdomain can have any arbitrary $r_{0k}$ as long as all the frequency 
lines in each $\Omega_k$ sum to the same global value. This flexibility is one of the
most important features of our algorithm.
	\item[\bf 4.] {\bf In each $\Omega_k$ an event is chosen with probability $p_{ik} = {r_{ik}}/{R_{max}}$, 
including null events with $p_{k0} = {r_{k0}}/{R_{max}}$.} For this step, we must ensure that independent 
sequences of random numbers be produced for each $\Omega_k$, using appropriate parallel pseudo random number generators.
	\item[\bf 5.] {\bf As in standard BKL, a time increment is sampled from an exponential distribution:
	\begin{displaymath}
    	\label{eq:step5}
    		\delta t_p = -\frac{\ln\xi}{R_{max}}
	\end{displaymath}
where $\xi\in(0,1)$ is a suitable random number}. 
Here, by virtue of Poisson statistics, $\delta t_p$ becomes the global 
time step for all of the parallel processes.
	\item[\bf 6.] {\bf Communicate boundary events.} A global call will always achieve communication
of boundary information. However, depending on the characteristics of the problem at hand, local calls may suffice,
typically enhancing performance.
\end{enumerate}

\subsection{The sublattice method for solving boundary conflicts}
\label{subsec:boundary}
\begin{figure}[h]
\begin{center}
  \includegraphics[width=6cm]{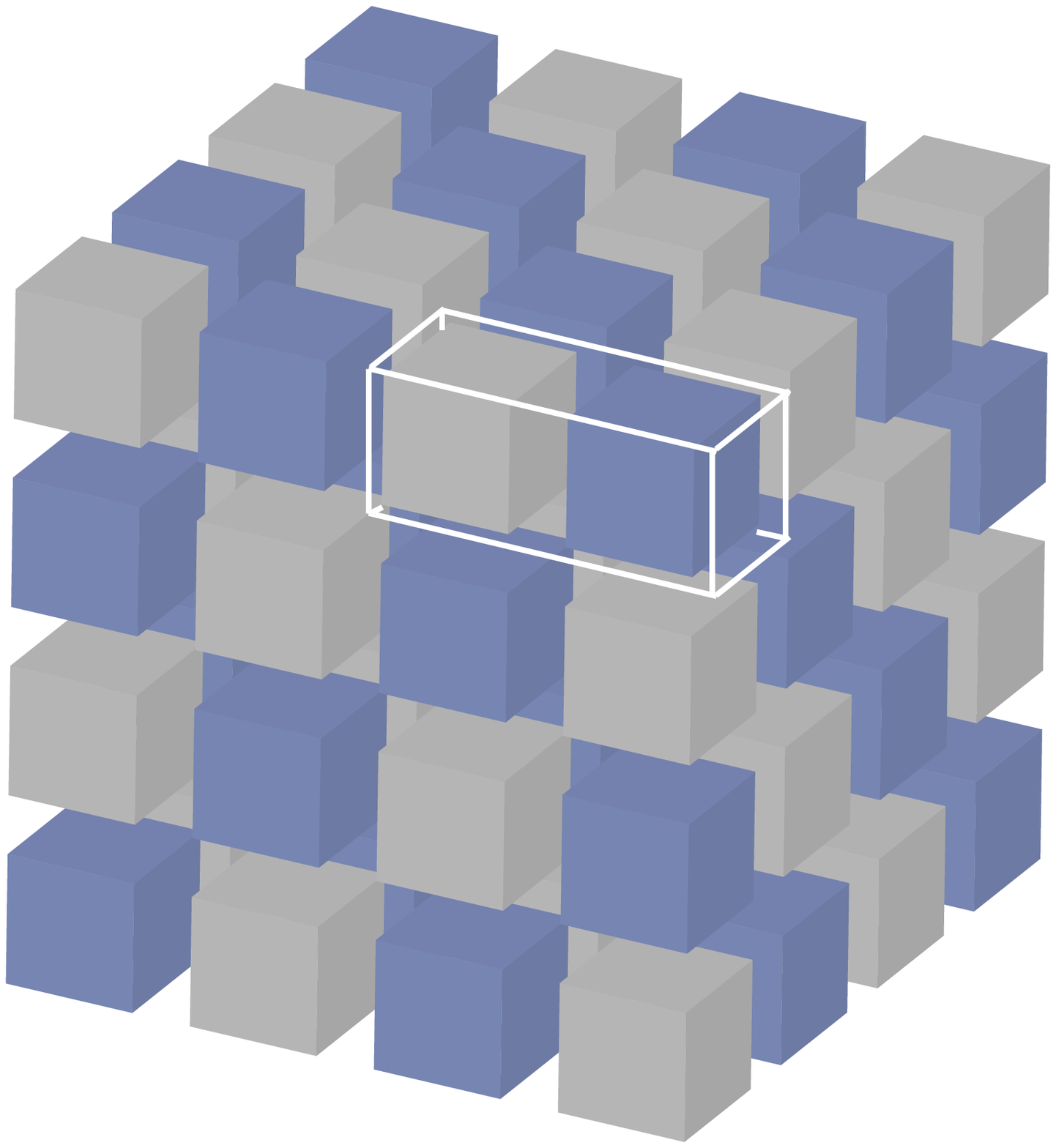}
  \includegraphics[width=6cm]{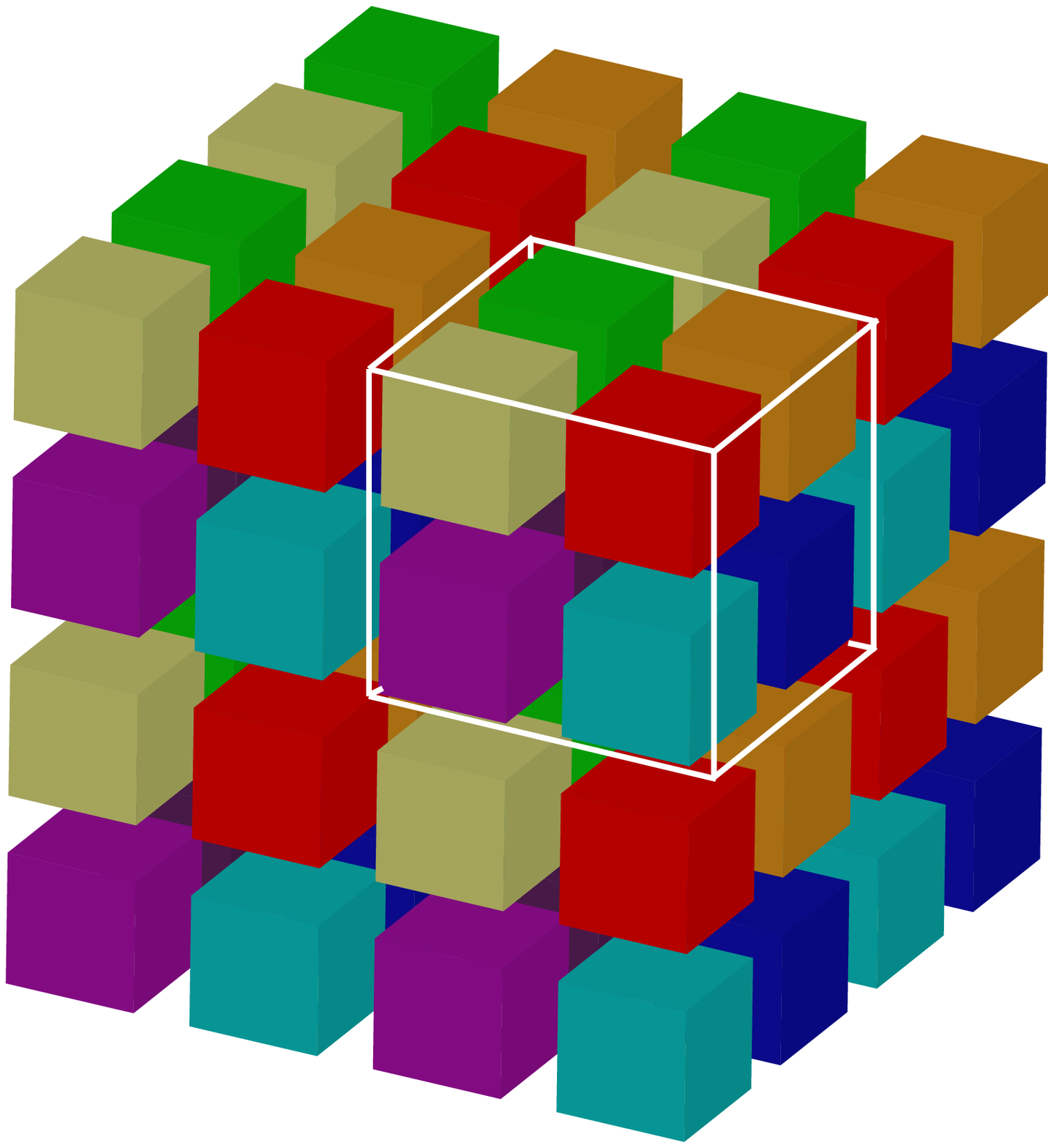}  
  \caption{Sublattice coloring scheme in three dimensions with regular subdivisions.
  Both two and eight color subdivisions are shown, corresponding to first and farther
  nearest neighbor sublattice interactions. The white wireframe indicates the
  indivisible color block assigned to processors}\label{fig:coloring}
\end{center}
\end{figure}

As we have shown, this algorithm solves the master equation exactly for 
non-interacting particles. When particles are allowed to interact across 
domain boundaries, suitable corrections must be implemented to avoid 
boundary conflicts. For lattice-based kinetics with short-ranged interaction 
distances this is straightforwardly achieved by methods based on the chessboard 
sublattice technique. This spatial subdivison method has been used in 
multispin calculations of the kinetic Ising model since the early 1990s 
\cite{hermann1991,oliveira1991,parisi1998}. In the context of parallel kMC algorithms,
Shim and Amar were the first to implement such procedure \cite{shim2005}, in which 
a sublattice decomposition was used to isolate interacting domains in each cycle. 
The minimum number of sublattices to ensure non-interacting adjacent domains 
depends on a number of factors, most notably system dimensionality
\footnote{In 2D, four sublattices are sufficient to resolve any arbitrary mapping, 
as established by the solution to the `four-color problem' \cite{appel1977}}. 
In 3D, the chessboard method requires 
a subdivison into a minimum of either two or eight sublattices, depending on whether
only first or farther nearest neighbor interactions are considered. This is schematically
shown in Figure \ref{fig:coloring}, where each sublattice is defined by a specific color.
The Figure shows the minimum sublattice block (white wireframe) to be assigend to each 
processor. These blocks are indivisible and each processor can be assigned only integer
multiples of them for accurate spkMC simulations.
The implementation of the sublattice algorithm is as follows.
Because here eq.\ \ref{eq:isingh} only involves first nearest neighbor interactions, 
the spatial decomposition performed in this work is such that it enables 
a regular sublattice construction with exactly two colors . 
In this fashion, each $\Omega_k$ becomes a subcell of a given sublattice, which imposes that each processor must 
have a multiple of two (or eight, for longer range interactions) number of subcells. 
Using a sublattice size greater than the particle interaction distance guarantees that no 
two particles in adjacent $\Omega_k$ interact. Step (4) above might then be substituted by the 
following procedure:

\begin{enumerate}
\item[{\bf 4a.}] {\bf A given sublattice is chosen for all subdomains}.  This choice may be performed in 
several ways such as fully random or using some type of color permutation so that every sublattice is 
visited in each kMC cycle. Here we have implemented the former, as, for example, in the  synchronous 
sublattice algorithm (SSL) of Shim and Amar \cite{shim2005}. The sublattice selection is performed 
with uniform probability thanks to the flexibility furnished by the spkMC algorithm, which takes advantage 
of the null rates to avoid global calls to communicate each sublattice's probability. Restricting each 
processor's sampling to only one lattice, however, while avoiding boundary conflicts, results in a systematic 
error associated with spatial correlations. The errors incurred by this procedure will be analyzed in 
Section \ref{sec:bias}.
\item[{\bf 4b.}] {\bf An event is chosen in the selected sublattice with the appropriate probability, including null events}. 
When the rate changes in each $\Omega_k$ after a kMC cycle are unpredictable, a global communication of $R_{max}$ in step (2) 
is unavoidable. When the cost of global communications becomes a considerable bottleneck in terms of parallel efficiency, 
it is worth considering other alternatives. For the Ising system, we consider the following:
\newline
\small
\begin{center}
\begin{boxedminipage}[l]{12cm}
\begin{itemize}
\item The simplest way to avoid global communications is to prescribe 
$R_{max}$ to a very large value so as to ensure that it is never surpassed 
regardless of the kinetics being simulated. For the Ising model, this 
amounts to {\it calculating the maximum theoretical aggregate rate for 
an ensemble of Ising spins}. For a given subdomain $\Omega_k$, this is:
$$ {R'}_{max} = \lambda n_k \left[\frac{\exp(-\Delta E_{max})}{1+\exp(-\Delta E_{max})}\right] $$
where $E_{max}$ is the theoretical maximum energy increment due to a 
single spin change:
$$ E_{max} = -2\left(n_b|J| - |H|\right)$$
and $n_b$ is the lattice coordination number. This procedure is very conservative 
and may result in a poor parallel performance.
\item {\it Perform a self-learning process to optimize ${R'}_{max}$}. This 
procedure is aimed at refining the upper estimate of $R_{max}$ by recording 
the history of rate changes over the course of a pkMC simulation. For example, one can 
start with the maximum theoretical aggregate Ising rate and start decreasing 
the upper bound to improve the efficiency. For this procedure, a tolerance 
to ensure that ${R'}_{max}>R_{max}$ must be prescribed. A sufficiently-long
time history of this comparison must be stored to perform regular checks and
ensure that the inequality holds. 
\end{itemize}
\end{boxedminipage}
\end{center}
\normalsize
\end{enumerate}
This algorithm solves the same master equation as the serial case but it is not strictly
rigorous. As we have pointed out, the sampling strategies adopted to solve boundary conflicts 
introduce spatial correlations that result in stochastic bias. 
Under certain conditions spkMC does behave rigorously in the sense that this 
bias is smaller than the intrinsic statistical error. We explore these 
issues in the following section. 

\section{Stochastic bias and analysis of errors}
\label{sec:bias}
The algorithm introduced above eliminates the occurrence of boundary conflicts 
at the expense of limiting the sampling configurational space of the system in 
each kMC step. Because boundary conflicts are inherently a spatial process, this 
introduces a spatial bias that must be quantified to understand the statistical 
validity of the spkMC results. 
Next, we analyze this bias by testing the behavior of the magnetization 
when the system is close to the critical point ($\sigma_c$).
All the results shown in this section correspond to the sublattice algorithm
using two colors with random selection.

\begin{figure}[h]
\begin{center}
\includegraphics[width=13cm]{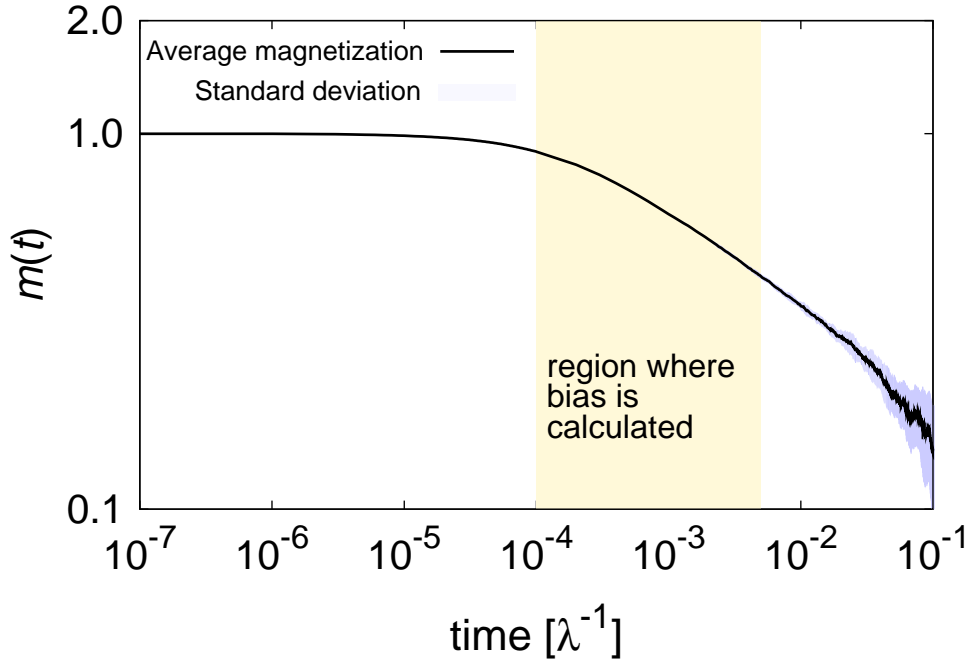}
\caption{Time evolution of the magnetization of a 262,144-spin Ising system
at the critical temperature. The curve is the result of 20 independent serial
kMC runs. The standard deviation, $\sigma_s$, is represented by the purple shaded area about
the magnetization curve. The golden shaded area marks the region over which the bias 
is computed.}
\label{fig:mag}
\end{center}
\end{figure}

The bias is defined as the difference between a parallel calculation and a reference 
calculation usually taken as the mean of a sufficient number of serial runs
\footnote{If available, an analytical solution may of course be used as a reference as well.}:
\begin{equation}
\mathrm{bias}=\langle m(\sigma_c)\rangle_p - \langle m(\sigma_c)\rangle_s
\label{eq:bias}
\end{equation}
where $\langle m(\sigma_c)\rangle_p$ and $\langle m(\sigma_c)\rangle_s$ are the 
averages of a number of independent runs in parallel and in serial, respectively, 
for a given total Ising system size. The initial ($m(t=0)=m_0$) and boundary (periodic) 
conditions in both cases are identical. In Figure \ref{fig:mag} we show the time
evolution of the magnetization of an $N$=262,144-spin ($2^{18}$) Ising system, averaged over 20 
serial runs, used as the reference for the calculation of the bias. The purple
shaded region gives the extent of the standard deviation, which is initially very small,
when $m_0$ is very close to one, but grows with time as the system approaches its 
paramagnetic state and fluctuations are magnified. The shaded region (in gold)
between $10^{-4}<t<5$$\times$$10^{-3}\lambda^{-1}$ has been chosen for convenience and 
marks the time interval over which eq.\ \ref{eq:bias} is solved\footnote{And where the critical exponent
in Section \ref{sec:app} is measured.}. 
The same exercise has been repeated for a 2,097,152-spin ($2^{21}$) sample
with 5 serial runs performed (not shown).

\begin{figure}[h]
  \begin{center}
    \includegraphics[width=14cm]{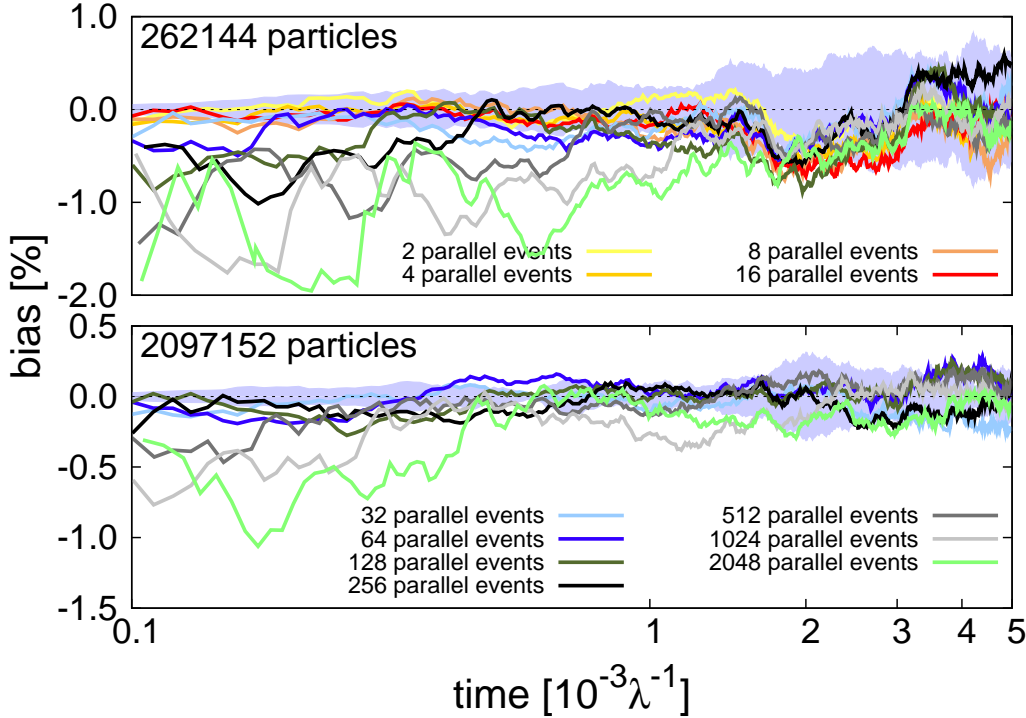}
    \caption{Time evolution of the parallel bias for two Ising system sizes at 
      the critical point. Results for parallel runs with several numbers of processors are 
      shown. The shaded region corresponds to the interval contained within
      the standard deviation of the reference (serial) case. Note the logarithmic scale for
      the abscissa.} 
    \label{fig:b-t}
  \end{center}
\end{figure}

Figure \ref{fig:b-t} shows the time evolution of the bias for a number of parallel
runs corresonding to the two system sizes studied. The shaded area in the figure corresponds 
to the interval contained within the standard deviation of the serial case (cf.\ Fig.\ \ref{fig:mag}). 
Therefore, this analysis yields the maximum number of parallel events that can be considered to obtain 
a solution statistically equivalent to that given by the serial case.

The figure shows up to what number of parallel processes can the serial and sublattice methods 
be considered statistically equivalent in the entire range where the bias is 
calculated. For the 262,144-spin system this is 32, whereas for the 2,097,152 one it is
approximately 256. However, runs whose errors are larger than the serial
standard deviation at short time scales ({\it e.g.} $\ge$64 and $\ge$512 for, respectively, 
the 262,144 and 2,097,152-spin systems) gradually reduce their bias as time progresses. 
In fact, at $t\gtrsim2\times10^{-3}\lambda^{-1}$, all parallel runs fall within $\sigma_s$.
It appears, therefore, that fluctuations play an 
important a role in the parallel runs for low numbers of processes an spkMC cycles. 
As the accumulated statistics increases (more cycles), this effect gradually disappears. 
In any event, the bias is never larger than $\approx$2\% for all cases considered here.

\begin{figure}[h]
\begin{center}
  \includegraphics[width=13cm]{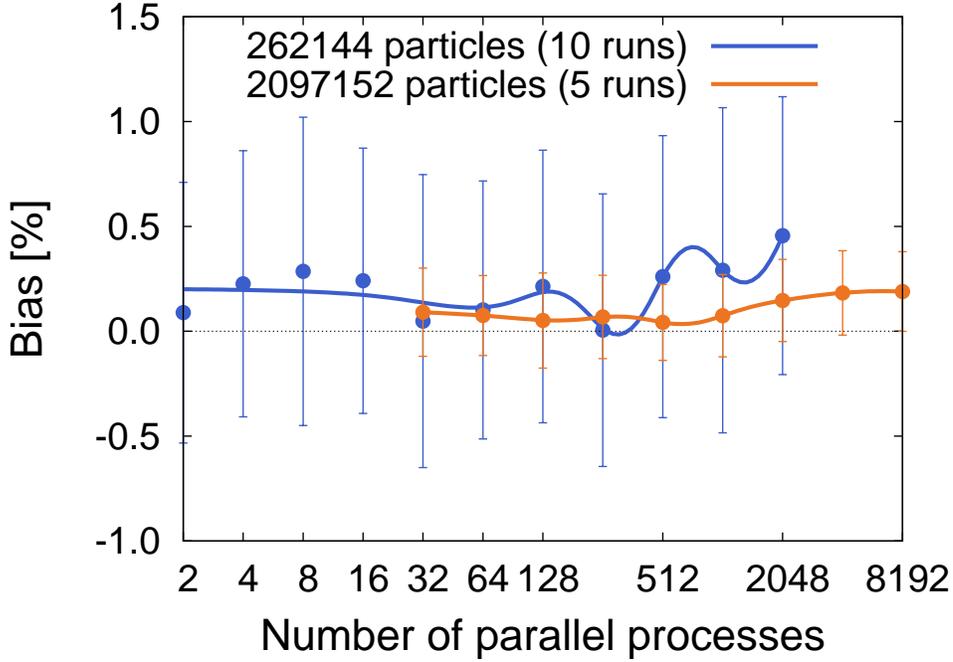}
  \end{center}
  \caption{Systematic parallel bias for 262,144 and 2,097,152-spin 3D Ising systems 
(obtained, respectively, from 10 and 5 independent runs) as a function of the number of parallel 
processes. Note that the number of parallel 
processes is equal to the total number of subcells divided by the number of different sublattices (=2, in 
our case).} 
  \label{fig:bias}
\end{figure}

Although Fig.\ \ref{fig:b-t}) provides an informative quantification of the
errors introduced by the parallel method, it is also important 
to separate this systematic bias from the statistical errors 
associated with each set of independent parallel runs. This is quantified by the standard 
deviation of the \emph{time-integrated} bias, defined as:
\begin{equation}
\sigma_b = \sqrt{\sigma_p^2 + \sigma_s^2}
\label{eq:dev}
\end{equation}
where the terms inside the square root are the parallel and serial variances respectively.
We next solve eqs.\ (\ref{eq:bias}) and  (\ref{eq:dev}) during the time interval prescribed
above for the two system sizes considered in Fig.\ \ref{fig:b-t}.
The absolute value of the systematic bias is extracted from a 
number of independent runs (10 and 5 respectively) and plotted in Figure \ref{fig:bias} 
as a function of the number of parallel processes. Note that the number of parallel 
processes is equal to the total number of subcells divided by the number of different sublattices 
(=2, in our case). 

The figure shows that the absolute value of the bias is always smaller than the 
statistical error ({\it i.e.} the error bars always encompass zero bias). 
This implies that, in the range explored, a given problem may be solved in parallel and 
the result obtained can be considered statistically equivalent to a serial run.
The bias is roughly constant and always below $0.5\%$ in the entire range explored for
both cases. However, the bias is consistently lower for the larger system size, as are 
the error bars. This is simply related to the moderation of fluctuations with system size.

An analysis such as that shown in Figure \ref{fig:bias} allows the user to control the 
parallel error by choosing the problem size and the desired number of particles per 
subcell. Consequently, our method continues to be a controlled approximation in the sense
that the error can be intrinsically computed and arbitrarily reduced.

\section{Algorithm performance}
\label{sec:eff}

\begin{figure}[h]
\begin{center}
\includegraphics[width=13cm]{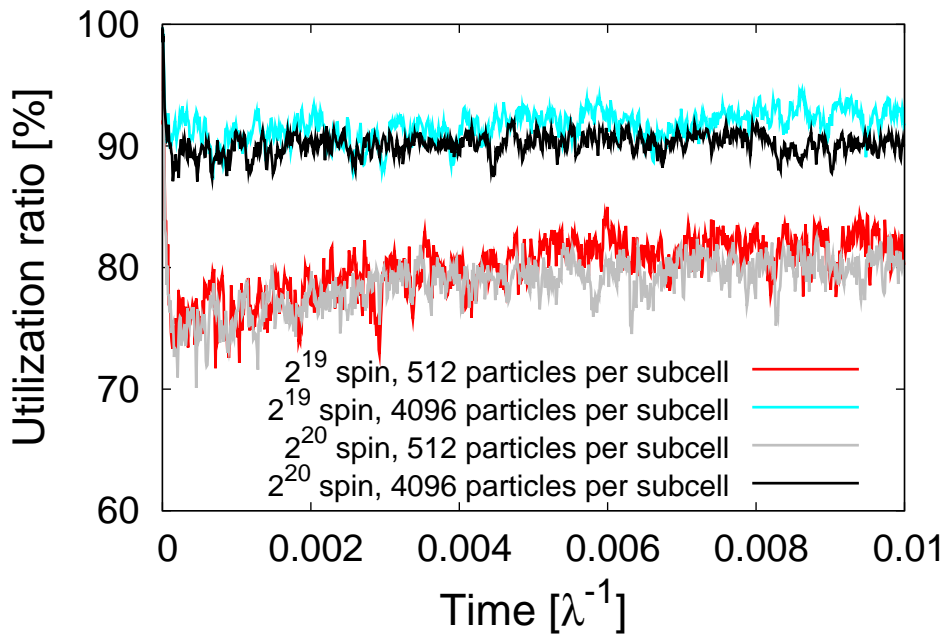}
\caption{Evolution of the utilization ratio (UR) with simulated time for 524,288 ($2^{19}$) and
1,048,576 ($2^{20}$) particle Ising system. Calculations have been done varying the number of sublattices
per processor, which is shown to have a significant impact on the UR.}
\label{fig:ur}
\end{center}
\end{figure}

The algorithm's performance can be assessed via its two fundamental contributions, namely, one that is 
directly related to the implementation of the minimal process method (MPM) through the null events 
\cite{hanusse}, and the parallel performance {\it per se}. The effect of the null events is quantified 
by the utilization ratio (UR):
\begin{equation}
\mathrm{UR}=1-\frac{\sum_k r_{0k}}{KR_{max}}
\label{eq:ur}
\end{equation}
which gives the relative weight of null events on the overall frequency line. 
The UR determines the true time step gain associated with the implementation of the MPM as \cite{martinez2008}:
$$\delta t^{*}=K\cdot\mathrm{UR}\cdot\delta t_s$$
where $\delta t^{*}$ and $\delta t_s$ are, respectively, the MPM and standard time steps. 
This procedure is intrinsically serial, and will result in superlinear scalar behavior if not 
taken into account for parallel performance purposes. 
Next, we show in Figure \ref{fig:ur} the evolution of the UR for 524,288 ($2^{19}$) and 1,048,576 
($2^{20}$) spin systems. We have done calculations for several numbers of processors and number of 
particles per subcell. We find that the determining parameter is the latter, {\it i.e.} for a fixed 
system size and number of processors used, the UR displays a strong dependence with the number of 
particles per subcell. The figure shows results for 512 and 4096 particles per subcell, which in 
the 524,288(1,048,576)-spin system amounts to, respectively, 1024(2048) and 128(256) subcells per 
processor. In the latter case, the UR eventually oscillates around $\sim82\%$, whereas in the former 
it is approximately 90\% ({\it i.e.} on average, $\sim18\%$ and 10\% of events, respectively, are \emph{null} events). 

For its part, the parallel efficiency is defined as the wall clock time employed in a serial calculation relative to 
the wall clock time of a parallel calculation with $K$ processors involving a $K$-fold 
increase in the problem size:
\begin{equation}
\eta = \frac{t_s(1)}{t_p(K)}
\label{eq:pe0}
\end{equation}
The inverse of the efficiency gives the weak-scaling behavior of 
the algorithm. Due to the absence of fluctuations that exist in other parallel algorithms based
on intrinsically asynchronous kinetics (cf., {\it e.g.}, Ref.\ \cite{shim2005}), the ideal parallel efficiency of 
pkMC is always 100\%.

Let us now consider the efficiency for the following weak-scaling problem. Assuming that 
frequency line searches scale linearly with the number of walkers in our system, 
the serial time expended in simulating a system of $N$ spins to a total time $T$ is:
$$ t_s(1) = n_s\left(t_{exe}+{\cal O}(N)\right) $$
where $n_s$ is the number of cycles required to reach $T$, and $t_{exe}$ is the 
\emph{computation} time during each kMC cycle. For its part, the total parallel time
for the $K$-fold system is:
$$ t_p(K) = n_p\left(t_{exe}+{\cal O}(N) + t_c\right) $$
where $n_p$ is the counterpart of $n_s$ and $t_c=t_g+t_l$ is the communications 
overhead due to global and local calls. In the most general case, the efficiency is then:
\begin{equation} 
\eta = \frac{n_s\left(t_{exe}+{\cal O}(N)\right)}{n_p\left(t_{exe}+{\cal O}(N) + t_c\right)} 
\label{eq:initial}
\end{equation}
As mentioned in the paragraph above, when it is ensured that the serial algorithm also take 
advantage of the time step gain furnished by the minimal process method, the number of cycles 
to reach $T$ is the same in both cases, $n_s\equiv n_p$. The parallel efficiency then becomes:
\begin{equation}
\eta = \frac{t_{exe}+{\cal O}(N)}{t_{exe}+{\cal O}(N) + t_g + t_l}
\label{eq:pe1}
\end{equation} 
Next, by virtue of the $\log P$ model \cite{culler1996}, we assume that the cost of global communications is
${\cal O}\left(\log K\right)^b$, with $b$ a constant, while the local communication time, $t_l$, 
is independent of the number of processors used and scales with the problem size as $\sqrt{N}$ \cite{schwabe1995}. 
If we consider the execution time $t_{exe}$ negligible compared to the communication time, we have: 
\begin{equation}
\eta = \frac{c_0N}{c_0N+c_1\left(\log K\right)^b+c_2\sqrt{N}}\\
     = \frac{1}{\left(c_1/c_0N\right)\left(\log K\right)^b+\left(1+c_2/c_0\sqrt{N}\right)}
\label{eq:pe2.5}
\end{equation} 
where $c_0$, $c_1$ and $c_2$ are architecture-dependent constants
The final expression then reduces to:
\begin{equation}
\eta = \frac{1}{a\left(\log K\right)^b+c}
\label{eq:pe3}
\end{equation}
where $a$ and $c$ are an architecture and problem dependent constants.

In the case where $R_{max}$ is overdimensioned {\it a priori} to a prescribed tolerance 
TOL of  the \emph{true} value,  $R^*_{max}\approx R_N(1+\mathrm{TOL})$, then $t_g=0$ and eq.\ \ref{eq:initial} becomes:
\begin{equation}
\eta = \frac{t_{exe}+{\cal O}(N)}{\left(1+\mathrm{TOL}\right)\left(t_{exe}+{\cal O}(N) + t_l\right)}
\label{eq:pe4}
\end{equation}
stemming from the fact that now the ratio $n_s/n_p=\delta t_p/\delta t_s=R_{max}/R^*_{max}$.
Assuming again that $t_{exe}$ is negligible with respect to $t_l$ and the ${\cal O}(N)$ term for frequency line
searches, the expression for the efficiency takes the form:
\begin{equation}
\eta = \frac{cN}{\left(1+\mathrm{TOL}\right)\left(cN + t_l\right)}
     = \frac{1}{c\left(1+\mathrm{TOL}\right)}
\label{eq:pe5}
\end{equation}
where $c$ is the same as in eq.\ \ref{eq:pe3}. 

Combining eqs.\ (\ref{eq:pe3}) and (\ref{eq:pe5}), we arrive at the criterion to choose the 
optimum algorithm:
$$\mathrm{TOL} < \frac{a}{c}(\log K)^b$$
{\it i.e.} as long as the above inequality is satisfied, avoiding global calls by conservatively setting
$R_{max}$ at the beginning of the simulation results in a more efficient use of parallel resources. Note 
that, via the constants $a$, $b$, and $c$, this is problem and machine-dependent, and establishing these
with confidence may require considerable testing prior to engaging in production runs.

\begin{figure}[h]
\begin{center}
\includegraphics[width=12cm]{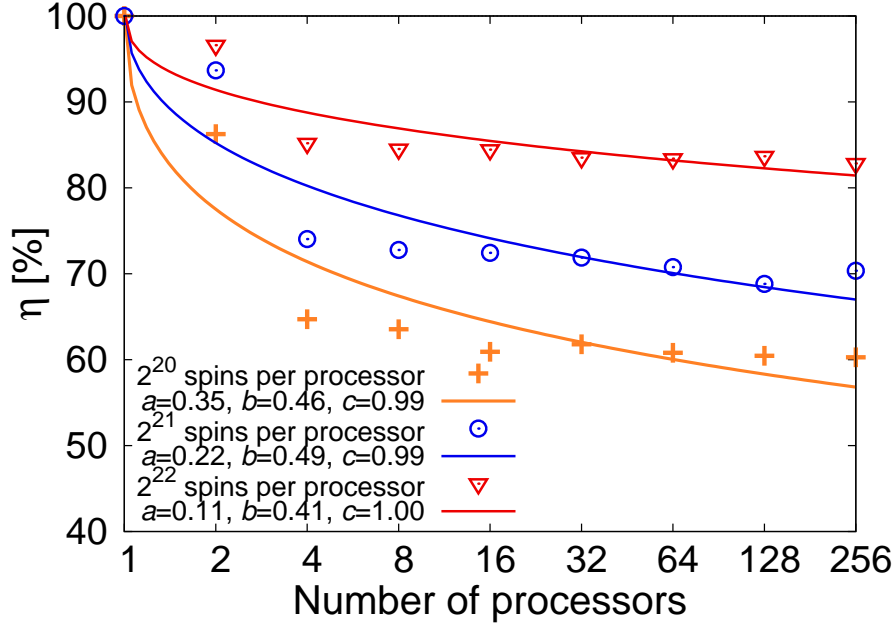}
\caption{Parallel efficiency for three different weak-scaling problems, one with $2^{20}$, 
one with $2^{21}$, and another with $2^{22}$ particles per processor of an Ising system at the 
critical point. The fitting constants $a$, $b$ and $c$ in eq.\ \ref{eq:pe3} are given for each case.} 
\label{fig:scaling}
\end{center}
\end{figure}

Next, we perform scalability tests for the case where $R_{max}$ is communicated globally, {\it i.e.} the 
efficiency is governed by eq.\ \ref{eq:pe3}. The tests have been carried out on LLNL's distributed-memory 
parallel platforms, specifically the ``hera'' cluster using Intel compilers \cite{mpi}. 
The scalabilitry calculations were all performed for 512 particles per subcell, regardless of the 
number of processors used, for systems with three different numbers of spins per processor, 
namely  4,194,304 ($2^{21}$), 2,097,152 ($2^{20}$), and 1,048,576 ($2^{19}$). This means that as
the number of particles per processor is increased, more subcells are assigend to each processor.
Figure \ref{fig:scaling} shows the parallel efficiency of the pkMC algorithm as a function
of the number of processors used for three reference Ising systems at the critical point. The fitting 
constants $a$, $b$ and $c$ are given for each case. 
As the figure shows, the number of spins per processor has a significant impact on the parallel efficiency, 
with larger sizes resulting in better performances. The efficiency at $K=256$ is upwards of 80\% 
for the largest system, and $\approx$ 60\% for the smallest system size.
The leap in efficiency observed in all cases between 2 and 4 processors is caused
by the nodal interconnects (band width) connecting quad cores in the platforms used. 
As expected from eq.\ \ref{eq:pe2.5}, the fitting 
constant $a$ scales roughly as $N^{-1}$, while $b$ does not display a large variability and
takes value between 0.41 and 0.49. $c$ can be considered equal to one for all three cases 
within the least-squares error, which implies negligible local communication costs 
($c_2\approx0$ in eq.\ \ref{eq:pe2.5}), 
and simplifies the tolerance criterion above to $\mathrm{TOL}<a(\log K)^b$.

The idea behind using a tolerance to minimize or contain global calls forms the basis of the so-called 
\emph{optimistic} algorithms \cite{merrick2007,kara2009}, where the parameter(s) controlling the
parallel evolution of the simulation are set conservatively ---either by a self-learning procedure or by
accepting some degree of error--- and monitored sparingly. For example, for the $2^{20}$-spin Ising
system, $a=0.11$, $b=0.41$, $c=1.00$, TOL varies between $<$0.09 and $<$0.22 in the range $2<K<256$. A
value of 0.09 may not be sufficient to encompass the time fluctuations in $R_{max}$, but it is expected that 
for a higher number of processors the efficiency will improve, although at the cost of the UR.
These and more aspects about the parallel efficiency and its behavior will be discussed in Section \ref{sec:conc}. 

\section{Application: billion-atom Ising systems at the critical point}
\label{sec:app}

\begin{figure}[h]
\begin{center}
\includegraphics[width=12cm]{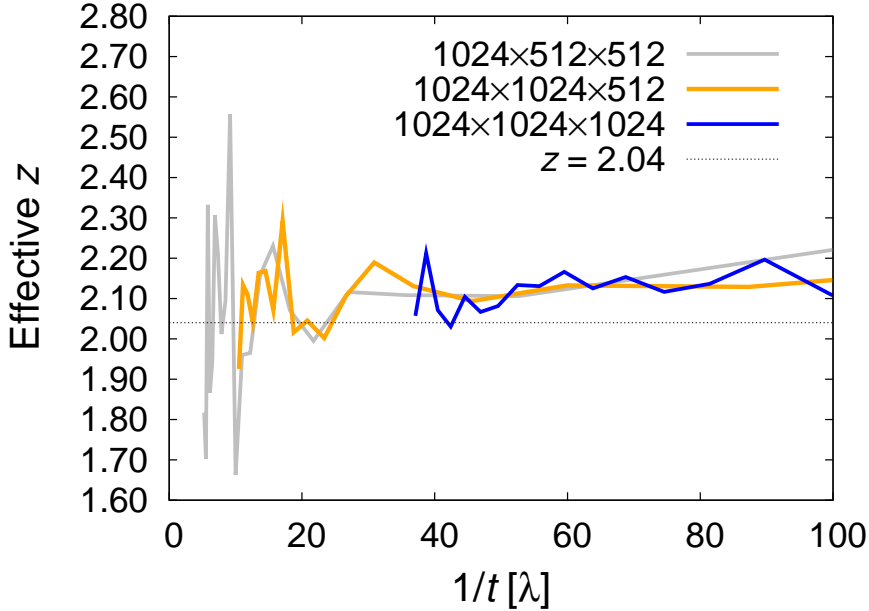}
\caption{Effective dynamical critical exponent as a 
function of inverse time using 1024
processors and 1,048,576 spins per subcell for approximately a quarter ($2^{28}$), 
half ($2^{29}$), and one ($2^{30}$) billion-spin Ising systems for $t>0.025\lambda^{-1}$. 
The horizontal line at $z=2.04$ marks the consensus value in 3D from the literature.}
\label{fig:z2}
\end{center}
\end{figure}

We now apply the method to study the time relaxation of large Ising systems near the critical point.
As anticipated in Section \ref{sec:Ising}, at the critical point, the relaxation 
time $\tau$ diverges as $\xi^z$, where $\xi\propto|T-T_c|^{-\nu}$. The scaling
at $T=T_c$ is then:
\begin{equation}
m(t)\propto t^{-\kappa/z\nu}
\end{equation}
In 3D, we use the known critical temperature $J/kT_c=0.2216546$ \cite{baillie1992,novotny2003} to find $z$.
We start with all spins $+1$ and let $m(t)$ decay from its initial value of unity 
down to zero. At each time point, we can find the critical exponent $z$ from:
\begin{equation}
z=-\frac{\beta}{\nu}\Big[\frac{d(\log m)}{d(\log t)}\Big]^{-1}
\end{equation}
where the ratio $\beta/\nu$ takes the known value of $0.515$ \cite{berche2004}.
We have carried out simulations with lattices containing 1024$\times$512$\times$512 ($2^{28}$), 
1024$\times$1024$\times$512 ($2^{29}$), and 1024$\times$1024$\times$1024 ($2^{30}$) spins. 
The results are shown in Figure \ref{fig:z2} for critical exponents calculated during $t>0.025\lambda^{-1}$,  
from time derivatives averaged over 300 to 500 timsteps.

At long time scales, the critical exponent oscillates around values that range from, roughly,
2.06 to 2.10 depending on system size. This in good agreement with the converged consensus value of
$\sim$2.04 published in the literature \cite{ivaneyko2006} (shown for reference in Fig.\ \ref{fig:z2}). However, 
as time increases, the oscillations increase their amplitude with inverse system size. 
Oscillations of this nature also appear in multi-spin calculations, both for smaller 
\cite{matz1994,grassberger1995,zheng2000} and larger \cite{stauffer1997} systems,
where the inverse proportionality with system size is also observed.
These may be caused by insufficient statistics due to size limitations, 
as we have shown that, under the conditions chosen for the simulations, 
our calculations are statistically equivelent to serial ones (cf.\ Fig.\ \ref{fig:bias}). 
The effect of the system size is also clearly manifested in the relative convergence rate
of $z$. As size increases, convergence to the expected value of 2.04 is achieved on much shorter 
time scales, {\it i.e.} fewer kMC cycles. 

\section{Discussion}
\label{sec:conc}

We now discuss the main characteristics of our method. We start by considering the 
three factors that affect the performance of our algorithm:
\begin{enumerate}[(i)]
\item \emph{Number of particles per subcell}. 
This is the most important variable affecting the algorithm's performance, as it controls
the intrinsic parallel bias and the utiliation ratio. Higher numbers of spins per subcell
both reduce the bias (cf.\ Figs.\ \ref{fig:b-t} and \ref{fig:bias}) and increase the UR
(Fig. \ref{fig:ur}), bolstering performance. However, this also results in an increase of the value 
of $R_{max}$, which causes a reduction in $\delta t_p$. Thus, decreasing the bias and 
increasing the time step are actions that may work in opposite directions in terms of performance, 
and a suitable balance between both should be found for each class of problems.
\item \emph{Number of particles per processor}. 
This parameter affects the  parallel efficiency via the number of spins per processor $N_k$ 
(for regular space decompositions, $KN_k\approx N$). As $N_k$ increases, a significant improvement 
is observed. This is directly related to the parameter $a$ in eq.\
\ref{eq:pe3}, which scales inversely with $N_k$ and is related to the cost of linear searches.
\item \emph{Total system size}. 
As Figs.\ \ref{fig:b-t} and \ref{fig:bias} show, for a given sublattice decomposition, 
a larger system incurs in smaller relative fluctuations 
in the magnetization, which results in a more contained bias. 
\end{enumerate}
Through the constants $a$, $b$, and $c$, the parallel efficiency strongly depends on the 
latency and bandwidth of the communication network used. That is why we have explored 
other efficiency-increasing alternatives that contain the number of global calls and 
the associated overhead. Prescribing a tolerance on the expected fluctuations of
$R_{max}$ is in the spirit of so-called optimistic kMC methods, and ideally its value is set by way of a 
self-learning procedure that maximizes the efficiency.

In any case, the intersection of items (i)-(iii) above configures the \emph{operational} 
space that determines the class of problems that our method is best suited for:
large (multimillion) systems, with preferrably a sublattice division that achieves an optimum 
compromise between time step gain and lowest possible bias, with the maximum possible number 
of particles per processor. These are precisely the conditions 
under which we have simulated critical dynamics of 3D Ising systems, with very good results. 
We conclude that 
spkMC is best designed to study this class of dynamic problems where fluctuations are important and there is 
unequivocal size scaling. This includes applications on many other areas of physics, such as crystal growth, 
irradiation damage, plasticity, biological systems, etc., although other difficulties due to the 
distinctiveness of each problem may arise that may not be directly treatable with the algorithm presented here.
We note that, because the parallel bias is seen to saturate  for large numbers of parallel 
processes, and the efficiency is governed by the inverse logarithmic term, the only limitation
to using spkMC is given by the number of available processors.

\section{Conclusions}
\label{sec:pep}
We have developed an extension of the synchronous parallel kMC algorithm presented in Ref.\ \cite{martinez2008}
to discrete lattices. We use the chess sublattice technique to rigorously account for boundary conflicts, and
have quantified the resulting spatial bias. The algorithm displays a robust scaling, governed by the global 
communications cost as well as by the spatial decomposition adopted.
We have applied the method to multimillion-atom three-dimensional Ising systems close to
the critical point, with very good agreement with published state-of-the-art results.
 
We thank M.~H.~Kalos for his invaluable guidance, suggestions, and inspiration, 
as this work would not have been possible without him.
This work performed under the auspices of the US Department of Energy by Lawrence Livermore National 
Laboratory under contract DE-AC52-07NA27344.
E. M. acknowledges support from the Spanish Ministry of Science and Education under the ``Juan de la Cierva''
programme.

\end{document}